# Polygonal Maxwell's fisheye lens via transformation optics as multimode waveguide crossing

S. Hadi Badri[a], H. Rasooli Saghai[b], Hadi Soofi[c]

[a] Department of Electrical Engineering, Azarshahr Branch, Islamic Azad University, Azarshahr, Iran

[b] Department of Electrical Engineering, Tabriz Branch, Islamic Azad University, Tabriz, Iran

[c] School of Engineering- Emerging Technologies, University of Tabriz, Tabriz 5166616471, Iran

sh.badri@iaut.ac.ir

## Abstract

Multimode waveguide crossings are crucial components for novel mode-division-multiplexing systems. One of the challenges of multimode waveguide routing in MDM systems is decreasing the inter-mode crosstalk and mode leakage of waveguide crossings. In this work, we present the intersections of three and four waveguides based on polygonal Maxwell's fisheye lens via transformation optics. The designed lenses are implemented by mapping their refractive index to the thickness of guiding Si layer. The three-dimensional finite-difference time-domain simulations are used to evaluate the performance of the proposed 3×3 and 4×4 crossings. The footprint of the 3×3 and 4×4 waveguide star crossings are 18.6×18.6 and 27.5×27.5 μm$^2$, respectively. For both waveguide crossings, the intermodal crosstalk in the output port is lower than -22dB while the crosstalk to other ports is lower than -37dB for TE$_0$, TE$_1$, and TE$_2$ modes. The insertion losses for these modes are lower than 0.5dB in a bandwidth of 415nm covering the whole optical telecommunication bands.

## Introduction

The ultimate goal of the nanophotonics is to squeeze a large number of optical components onto a single chip. Hence, crossing of waveguides connecting these components is inevitable. Various methods for designing a broadband waveguide crossing with low insertion loss and low crosstalk levels have been proposed. Most notable photonic crystal waveguide crossings include designs based on resonant cavity [1-3], coupled-cavity waveguide [4], utilizing the symmetric properties of the propagation modes of square-lattice [5], nonidentical coupled resonator waveguides [6], cascading cavities [7, 8], topology optimization [9], Wannier basis design and optimization [10], and self-collimation phenomenon [11]. The intersections based on resonant cavities have inherently narrow bandwidth with crosstalk levels below -30*dB*. To increase the bandwidth, the Q-factor is decreased, resulting in weaker mode-matching between the waveguides and resonant cavities and consequently lower transmission. There has been no report of multimode waveguide crossing based on the above methods [1-11] and they only support a single propagating mode. Silicon-on-insulator (SOI) waveguide crossings can be designed based on multimode interference (MMI) [12-18], mode expanders [19, 20], subwavelength grating [21, 22], and wavefront matching [23, 24]. The designs based on MMI typically have bandwidth of 60-100*nm*, crosstalk lower than -18*dB*, and with footprints larger than 4.8μm×4.8μm. Expanding the MMI designs to support higher order modes is challenging due to the different self-imaging distances for each mode. On the other hand, designs based on mode expander have narrower bandwidth of 20-25*nm*, desirable crosstalk levels i.e., lower than -40*dB*, and footprint on the same order of MMI devices. Utilizing the wavefront matching technique, crossing angles can be scaled down to 5° with footprints as large as 120μm×230μm. Single-mode waveguide crossing based on linear tapers with etched lens-like structure with a footprint of 1μm×1μm, 40*nm* bandwidth, and crosstalk below -30*dB* has been demonstrated [25]. Inverse-designed single-mode waveguide crossings such as 4×4 crossing with 5.3μm×5.3μm footprint, 80*nm* bandwidth, and crosstalk levels below -22.5*dB* [26], and polarization-insensitive waveguide crossing with 3μm×3μm footprint, 200nm bandwidth, and crosstalk levels below -28dB [27] have been proposed. However, transmission of higher-order modes has not been studied in these designs [25-27]. Multimode waveguide crossings can be designed by employing subwavelength asymmetric Y-junctions with 80*nm* bandwidth, 34μm×34μm footprint, and crosstalk lower than -24*dB* [28]. Because of the limited splitting ability of the Y-junction, it only supports two modes. Recently, designing single-mode or multimode waveguide crossings based on the Maxwell's fisheye lens (MFE) lens has attracted a considerable attention [29-31]. A waveguide crossing of four multimode waveguides has been designed with transformation optics (TO) based on MFE lens with a bandwidth of 390nm and crosstalk of less than -20dB [32]. The fabricated sample of this design has a footprint of 42μm×42μm. The results of [30, 32] prove that the abberation-free imaging properties of the MFE lens can be exploited to design waveguide crossings.

In this article, 3×3 and 4×4 multimode waveguide crossings are theoretically designed and numerically evaluated. The presented devices exhibits a bandwidth of 415nm (from 1260 to 1675nm) with crosstalk levels lower than -37dB. The imaging property of the MFE lens makes it possible to image the incoming guiding modes to the edge of the lens to the diagonally opposite side of the lens. The wavefront of the waveguides and the circular MFE lens does not match resulting in losses for higher order modes [32]. Transformation optics (TO) enables us to transform an optical component with a known geometry into infinite number of geometries with an identical optical response [33-35]. We utilized TO technique to provide a flat wavefront at the boundary of the lens, by transforming the circular MFE lens into the hexagonal and octagonal MFE lenses. The MFE lens is a gradient index (GRIN) medium with a refractive index defined as [31]

$$n_{lens}(r) = \frac{2 \times n_{min}}{1+(r/R_{lens})^2} \quad , \quad (0 \leq r \leq R_{lens}) \tag{1}$$

where $n_{min}$ is the minimum refractive index of the lens at its edge, $r$ is the radial distance from the center of the lens and $R_{lens}$ is the radius of the lens.

## 2. Design of polygonal Maxwell's fisheye lens

TO gives us a tool to transform any optical device with the given geometry into infinite number of new geometries with the same optical response. However, transforming a given coordinate space, virtual domain, into a new arbitrary one, physical domain, introduces some limitations such that the required permittivity and permeability may become anisotropic with extremely high or low values [36]. The resonant metamaterials used in implementation of these extreme values severely limit the bandwidth of the optical device. Provided that the transformed device includes only limited anisotropy and sub-unity refractive index regions, some simplifications have been proposed [37]. Nevertheless, designed material complexities can be minimized by quasi-conformal transformation optics (QCTO). In QCTO technique, angles between the coordinate lines in virtual and physical domains are maintained [38, 39]. Inherently two-dimensional QCTO is applied to design planar photonic components.

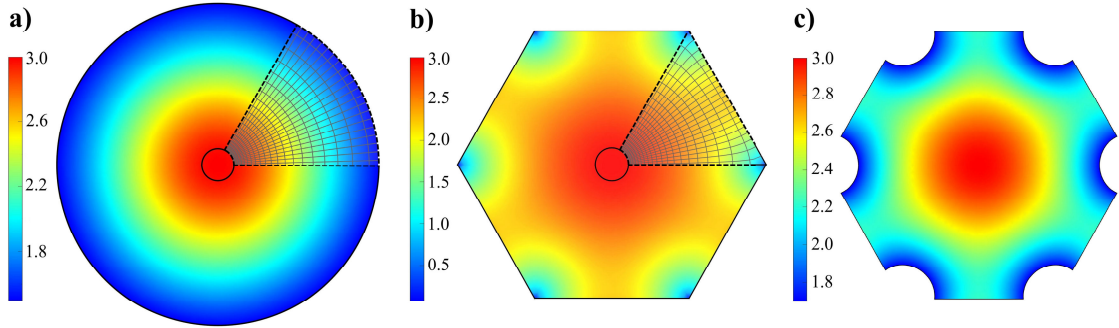

Figure 1. The refractive index, orthogonal grid, and quadrilateral domains of the a) virtual and b) physical. The quadrilateral domains are specified by dashed lines. c) Truncated hexagonal MFE lens used as waveguide crossing medium.

In this work, the circular MFE lens in the virtual domain is transformed to a polygon in the physical domain [40] . The first step in QCTO is to generate an orthogonal grid, i.e., grid lines are orthogonal to each other, in the virtual and physical domains [38]. The orthogonal grid is generated by solving the Laplace equation. Boundary orthogonality is achieved by applying Dirichlet-Neumann boundary conditions. Knowing that inverse of a conformal mapping is conformal, two domains with the same conformal module, M, can be mapped onto each other conformally by mapping them onto an intermediate domain. The intermediate domain is a rectangle with the same conformal module, M [41]. Conformal module is the ratio of the lengths of the two adjacent sides of a domain. Since only the refractive index and wavefront mismatches at the edge of the MFE lens with the waveguides are important, as shown in Figure 1, we excluded the inner center of the MFE lens with r<1μm from the transformation. For the 3×3 waveguide crossing, three waveguides intersect the MFE lens in six sides so the lens is divided into six equal parts. The quadrilateral virtual domain formed by this manner is displayed in Figure 1(a). The quadrilateral virtual domain is transformed to the quadrilateral physical domain of Figure 1(b). The dash lines specify the sides of the quadrilateral of virtual and physical domains used in QCTO. The generated orthogonal grid and the refractive indices of virtual and physical domains are also shown in this figure. The rectangular intermediate domain is not shown in this figure. Through our transformation, the circular edge is flattened and the refractive index of the lens at its edge is increased in the physical domain. Our simulations reveal that the light wave does not pass from the corners of the hexagonal lens. Therefore, we truncated the corners of the hexagonal MFE lens, to simplify the implementation of the lens, as displayed in Figure 1(c). The curves used in the truncation of the lens correspond to the refractive index contour level of 1.7. The material properties of the virtual domain are

$$\mu' = 1 \quad , \quad \varepsilon' = n_{vir}^2(r') = \begin{cases} n_{lens}^2(r') & r' \leq R_{lens} \\ n_{min}^2 & r' > R_{lens} \end{cases} \quad (2)$$

where $\varepsilon'$ and $\mu'$ are permittivity and permeability of the virtual domain. We have chosen $n_{min}$=1.45 and the diameter of the lens is 10μm. In our design, the intermediate domain happens to be the same as the physical domain. The transformation from virtual domain $(x', y', z')$ to physical domain $(x, y, z)$ is described with

$$x = x(x', y') \quad , \quad y = y(x', y') \quad , \quad z = z' \quad (3)$$

which is mapped to the material properties by

$$\varepsilon = \frac{A\varepsilon' A^T}{|A|} \quad , \quad \mu = \frac{A\mu' A^T}{|A|} \quad (4)$$

where $\varepsilon$ and $\mu$ are permittivity and permeability of the physical domain, respectively. $A$ is the Jacobian transformation matrix between the virtual and physical domains:

$$A = \begin{bmatrix} \frac{\partial x}{\partial x'} & \frac{\partial x}{\partial y'} & 0 \\ \frac{\partial y}{\partial x'} & \frac{\partial y}{\partial y'} & 0 \\ 0 & 0 & 1 \end{bmatrix} \quad (5)$$

when Cauchy-Riemann conditions are satisfied, i.e.

$$\frac{\partial x}{\partial x'} = \frac{\partial y}{\partial y'} \quad , \quad \frac{\partial y}{\partial x} = -\frac{\partial x}{\partial y'} \quad (6)$$

then the material properties in transverse electric (TE) mode, where the electric field is parallel to the z-axis, can be calculated by

$$\mu = 1 \quad , \quad \varepsilon = \frac{n_{vir}^2(r)}{|\det A|} \quad (7)$$

The method proposed to design a 3×3 waveguide crossing can be expanded to design a 4×4 waveguide crossing. To design a 4×4 waveguide crossing, we need to transform a circular MFE lens into an octagonal one. Similar to Figure 1(a), we divide a circle with a radius of 15μm into eight equal parts. Afterwards, a quadrilateral with curved side created with this method is transformed to a quadrilateral with flat side. The transformed quadrilateral is rotated with N×45 degrees where N=1, 2, 3, …,7. Eventually, the transformed octagonal MFE lens is obtained (Figure 2 (a)). Due to the negligible effect of the lens's corners on its performance, the lens is truncated as illustrated in Figure 2(b). The curves used in the truncation of the octagonal lens correspond to the refractive index contour level of 1.65.

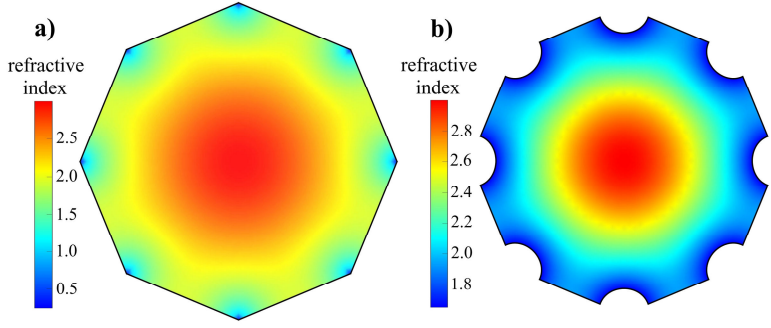

Figure 2. a) The octagonal MFE lens, b) Truncated octagonal MFE lens,

## 3. Implementation of the polygonal MFE lens

We have implemented the designed hexaxgonal and octagonal MFE lenses on SOI platform with varying the thickness of Si slab waveguide. GRIN lenses can also be implemented by graded photonic crystal (GPC) or multilayer structures [42]. The refractive index profile of the MFE lens can be implemented by varying the height of a silicon slab waveguide on top of a $SiO_2$ substrate. Air is considered as the top cladding material [32]. The effective index method (EIM) was applied to map the refractive index to the thickness of the Si layer. The silicon slab shown in Figure 3 with silicon dioxide substrate and air cladding was considered in the effective refractive index calculations. For the TE mode, where light propagates in the z direction, the electric field is [43]

$$E_y(x) = \begin{cases} Ce^{-qx} & x \geq 0 \\ C[\cos(hx) - \frac{q}{h}\sin(hx)] & 0 \geq x \geq -t \\ C[\cos(ht) + \frac{q}{h}\sin(ht)]e^{p(x+t)} & x \leq -t \end{cases} \quad (10)$$

and

$$q = \sqrt{\beta^2 - k_0^2 n_{Air}^2}, \quad p = \sqrt{\beta^2 - k_0^2 n_{SiO_2}^2}, \quad h = \sqrt{k_0^2 n_{Si}^2 - \beta^2} \quad (11)$$

where $k_0 = 2\pi/\lambda_0$ is the free-space wavenumber, $\beta = k_0 n_{eff}$ is the propagation constant, and $C$ is a constant. The eigenvalue equation for the TE modes of the slab waveguide is

$$\tan(ht) = \frac{p+q}{h(1 - \frac{pq}{h^2})} \quad (12)$$

where the only unknown quantity is $\beta$. The discrete values of $\beta$ satisfying Eq. 12 are the modes of the slab waveguide. The calculated values of $\beta$ are used to determine the $n_{eff}$ of that mode. The calculated $n_{eff}$ of $TE_0$ mode is shown in Figure 3. In the effective refractive index calculations, the thickness of 3μm was considered for silica substrate and air cladding.

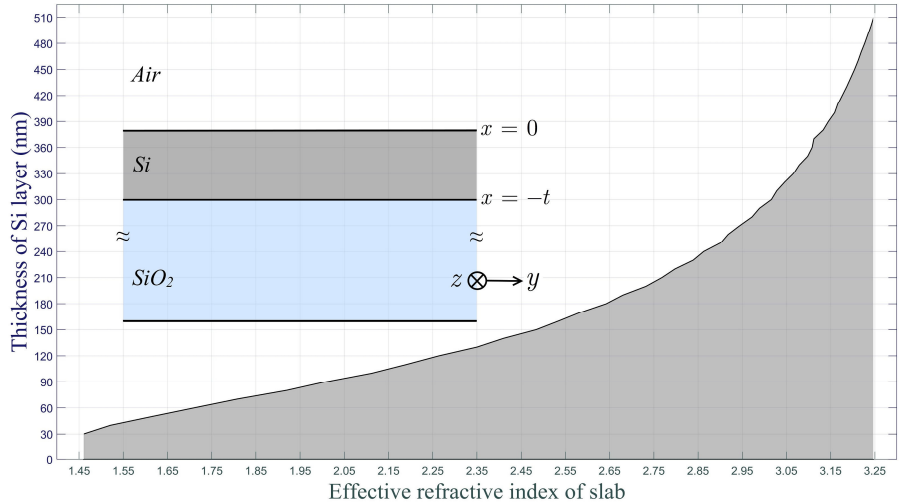

Figure 3. Silicon slab waveguide used for effective index calculation and mapping of silicon thickness to effective refractive index of slab waveguide are demonstrated.

The truncated hexagonal and octagonal MFE lenses implemented with this method are shown in Figure 4. The underlying SiO$_2$ layer and air cladding are not shown in this figure. The corners of the lenses were not implemented since the refractive index of the transformed lenses are lower than unity and their contribution to the performance of the crossing were negligible.

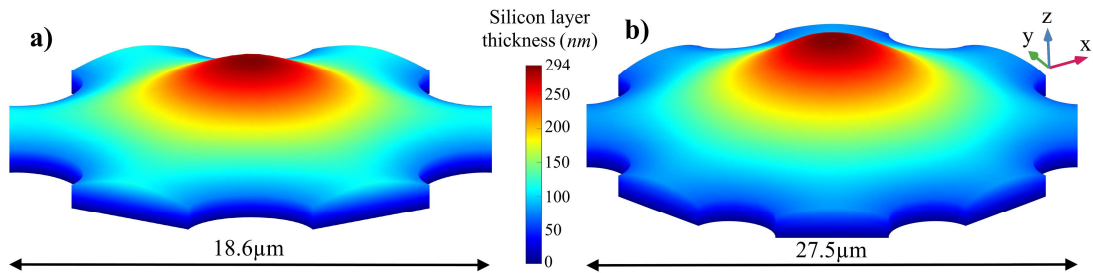

Figure 4. The implementation of a) hexagonal and b) octagonal MFE lenses based on varying the guiding layer thickness. The SiO$_2$ and air claddings are not shown in the 3D implementation.

## 4. Results and discussion

The three-dimensional (3D) finite-difference time-domain (FDTD) numerical simulations were carried out to evaluate the performance of the lenses implemented with varying the thickness of guiding layer as waveguide crossing. Figure 5 depicts the three crossing waveguides, the lens, and the power-streams of the TE$_1$ mode. In this figure, the underlying silica substrate and air upper cladding are not shown. The effective refractive index of the waveguides is considered as 2.2 and hence the thickness of the waveguides was chosen as 110nm. The width of the waveguides was chosen as 3μm to support at least three modes. It should be noted that the refractive index of the lens at the middle of its interface with the waveguides is 2.22 but it slightly decreases to 2.07 at the edges of waveguides. This translates into thickness variation in the interface of the waveguides and the lens which is obvious in Figure 5. However, our simulation results indicate that the slight refractive index variation of the lens at its interface with the waveguides has negligible effect on the performance of the crossing. Noticeable step-like changes in the thickness of the guiding layer would increase reflection. The gradual thickness variation of the guiding layer in our implementation ensures low reflection of the waveguide crossing. A constant refractive index in the interface of the lens and waveguides can be achieved by increasing the size of the lens. By easing this constraint, we have been able to reduce the footprint without degrading the performance of the crossing. For the 3×3 crossing, the average insertion losses are 0.14, 0.27, and 0.38 dB for the $TE_0$, $TE_1$, and $TE_2$ modes, respectively. Crosstalk levels at the ports of *in2*, *out2*, *in3*, *out3* are below -53, -46, and -43 dB for the $TE_0$, $TE_1$, and $TE_2$ modes, respectively. In addition, the intermodal crosstalk at the output port is below -23dB for these modes. The designed waveguide crossing has an ultra-wide bandwidth covering 1260-1675nm.

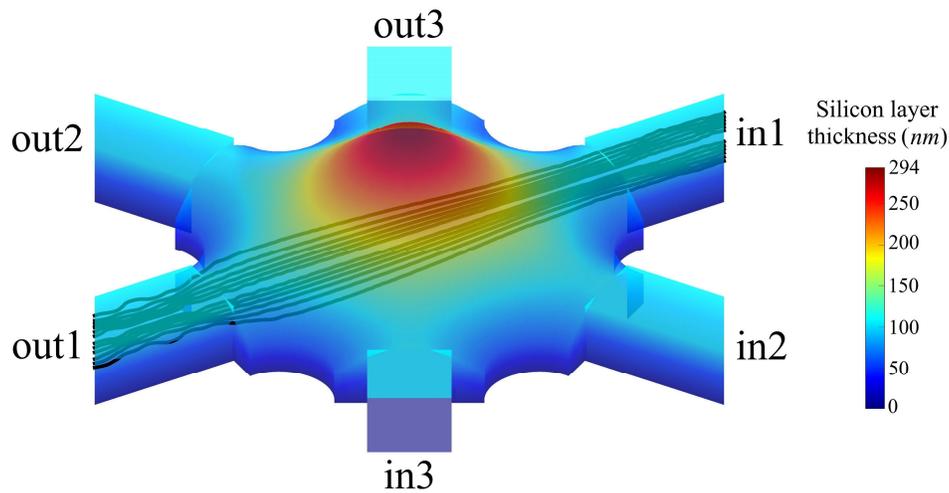

Figure 5. The 3×3 waveguide intersection based on the truncated hexagonal MFE lens. The power-streams illustrate the energy flow of the $TE_1$ mode. The silica substrate and air cladding are not shown in this figure.

For the 4×4 waveguide crossing, the thickness of Si layer in the waveguides was 80nm with a width of 3μm. The reason for this choice is that the thickness (or refractive index) of the octagonal lens at its edges is smaller than the hexagonal lens. This is apparent in Figure 4. The $H_z$ field distribution of the $TE_0$, $TE_1$, and $TE_2$ modes for a light of 1550nm wavelength are displayed in Figure 6. The light signal is injected from *in1* port. The reflection, transmission, and crosstalk at 1550nm are also shown in this figure. The scattering parameters of the $TE_0$, $TE_1$, and $TE_2$ modes are shown in Figure 7. The average insertion loss in the bandwidth of 1250-1675nm is 0.16, 0.21, and 0.33 for the $TE_0$, $TE_1$, and $TE_2$ modes, respectively. The intermodal crosstalk is lower than -22dB in the *out1* port while crosstalk levels at other ports are lower than -37dB.

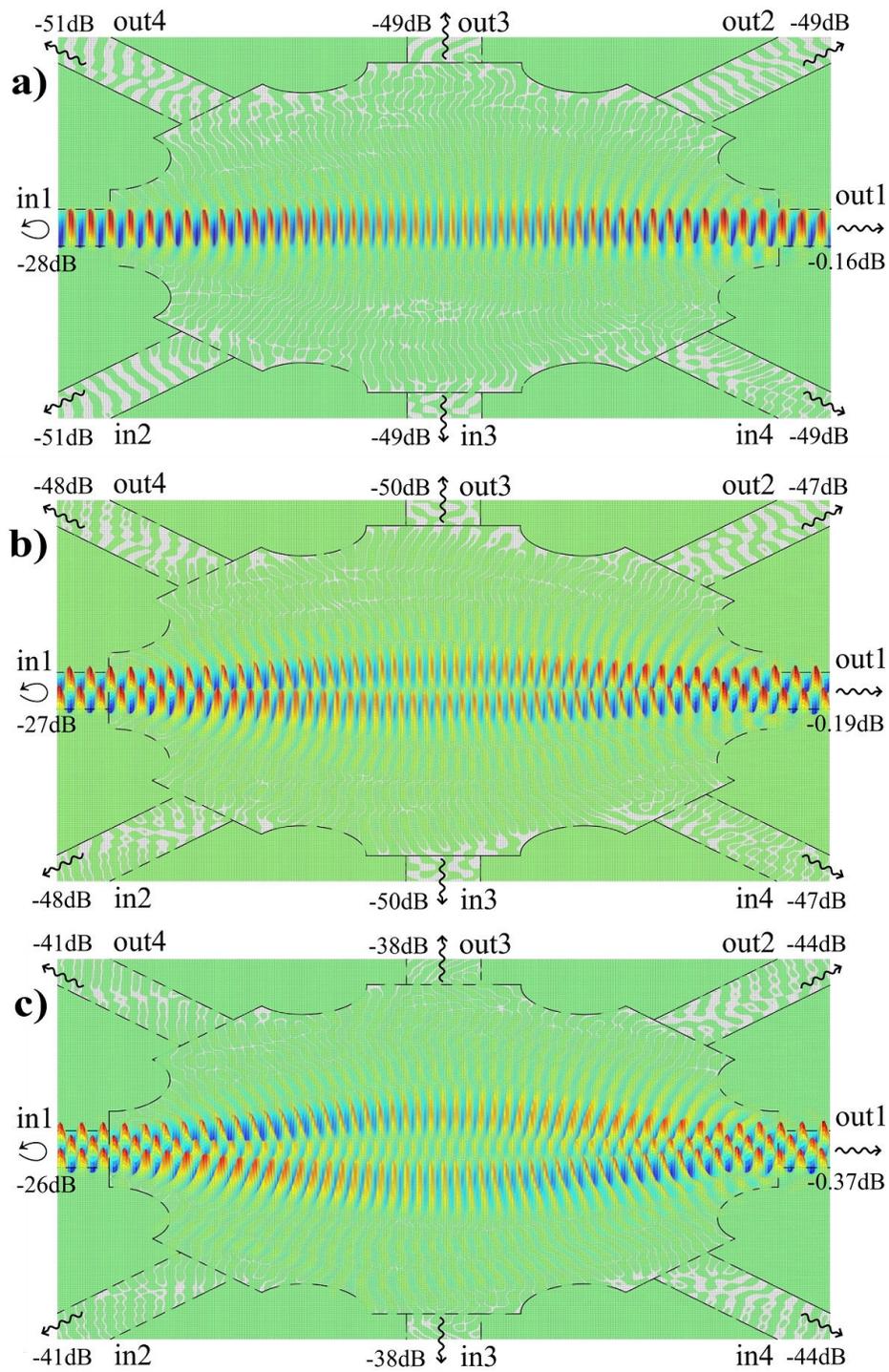

Figure 6. The $H_z$ field distribution for a) TE0, b) TE1, c) TE2 modes at 1550nm. The scattering parameters corresponding to 1550nm are also shown.

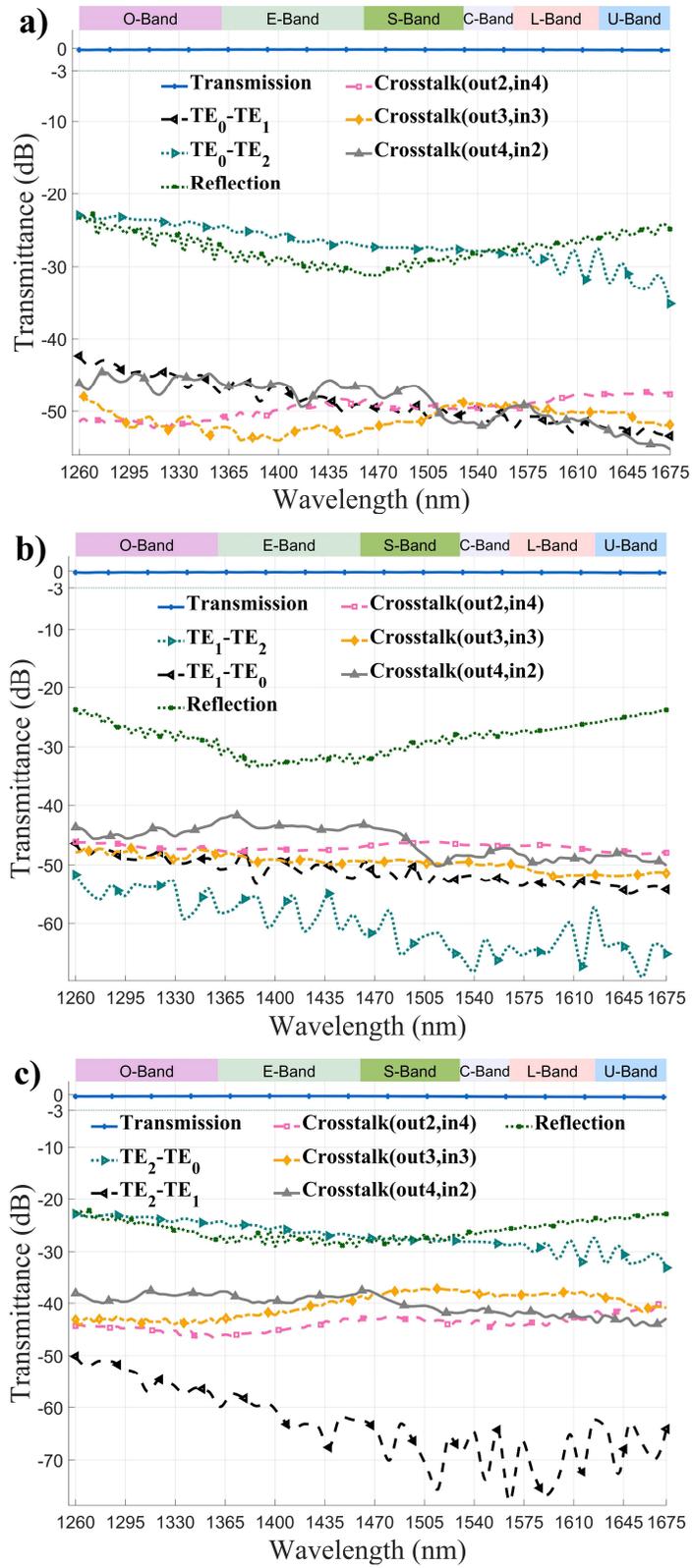

Figure 7. The scattering parameters for a) TE0, b) TE1, and c) TE2 modes.

## 4.1 Comparison with previous works

The characteristics of the designed multimode intersections and the references of [17, 18, 28, 30, 32] are summarized in Table 1. The crossing mechanism, insertion loss, central wavelength, bandwidth, crosstalk, footprint, number of supported modes, and number of crossing waveguides are compared in this table. Since the insertion loss usually increases as the order of the modes increases, the insertion loss of the highest-order mode supported by the crossing is reported in the table. References [17,18,28] only report 2×2 crossings so we focus on comparing our design with [30,32]. First of all, we should acknowledge that references [30,32] report experimental measurements while our results are based on numerical simulations. The insertion loss for this work, [30], and [32] are 0.33, 0.3, and 2.68dB, respectively. On the other hand, the simulation results of [32] predict the maximum insertion loss of 1.14dB while our design has the maximum insertion loss of 0.5dB. Due to the limitations in the measurement setup, [30] and [32] reported limited bandwidth but the simulation results of the [32] indicate that it has a bandwidth of 394nm. We report a 415nm bandwidth which is in the same range as [32]. However, reference [30] which is implemented by GPC, only reports the bandwidth of 1500-1600nm. Its performance may be degraded in lower wavelengths due to the step-wise profile of the GPC structure. The implementation of our work and [32] are based on varying the thickness of slab waveguide which imposes no bandwidth limitation to the device. In the 4×4 crossings, reference [30] has the smallest footprint, however, this was achieved by truncating the MFE lens. This may degrade the imaging properties of the lens which consequently increases the intermodal crosstalk. Reference [30] does not report the intermodal crosstalk. We took a simpler approach in the transformation of the MFE lens compared to [32] and we also relaxed the constraint of constant refractive index at the sides of the designed lens without introducing significant loss. By this method, we were able to reduce the 4×4 crossing's footprint by more than 50% compared to [32]. We have also incorporated the fabrication imperfections in our simulations by introducing random deviations to the designed lens. Our simulations show that the introduction of 15% random Si thickness deviations from the designed thickness has negligible effect on the insertion loss of the device.

*Table 1. Comparison of the proposed design and other multimode waveguide intersections*

| Ref. | Crossing Mechanism | Insertion Loss (dB) | $\lambda_{center}$ (nm) | Bandwidth (nm) | Cross-talk (dB) | Footprint μm² | number of supported modes | number of crossing waveguides |
|---|---|---|---|---|---|---|---|---|
| [17] | MMI coupler | 0.6 | 1560 | 60 | -24 | 4.8×4.8 | 2 | 2 |
| [18] | MMI coupler | 1.5 | 1560 | 80 | -18 | 30×30 | 2 | 2 |
| [28] | Asymmetric Y-Junction | 0.9 | 1560 | 80 | -24 | 34×34 | 3 | 2 |
| [30] | MFE lens | 0.3 | 1550 | 100 | -20 | 18×18 | 2 | 4 |
| [32] | MFE lens | 2.68 | 1550 | 40 | -33 | 42×42 | 3 | 4 |
| Octagonal lens | MFE lens | 0.33 | 1467 | 415 | -37 | 27.5×27.5 | 3 | 4 |

# 5. Conclusion

MDM can increase the bandwidth density in on-chip optical interconnects. Multimode waveguide crossing is one of the important building blocks in MDM systems. We designed the 3×3 and 4×4 multimode waveguide crossings with 18.6×18.6 and 27.5×27.5 μm$^2$ footprints, respectively. For the 4×4 crossing, the average insertion losses of 0.16, 0.20, and 0.33 dB for the $TE_0$, $TE_1$, and $TE_2$ modes are achieved, respectively. The intermodal crosstalk is below -22dB at the output port while the crosstalk levels at other ports are lower than -43dB. The proposed waveguide crossing covers the entire O, E, S, C, L, and U bands of optical communication.

# References

[1] Y.-G. Roh, S. Yoon, H. Jeon, S.-H. Han, Q.-H. Park, Experimental verification of cross talk reduction in photonic crystal waveguide crossings, Applied physics letters, 85 (2004) 3351-3353.
[2] S.G. Johnson, C. Manolatou, S. Fan, P.R. Villeneuve, J. Joannopoulos, H. Haus, Elimination of cross talk in waveguide intersections, Optics Letters, 23 (1998) 1855-1857.
[3] Y. Yu, M. Heuck, S. Ek, N. Kuznetsova, K. Yvind, J. Mørk, Experimental demonstration of a four-port photonic crystal cross-waveguide structure, Applied Physics Letters, 101 (2012) 251113.
[4] S. Lan, H. Ishikawa, Broadband waveguide intersections with low cross talk in photonic crystal circuits, Optics letters, 27 (2002) 1567-1569.
[5] Z. Li, H. Chen, J. Chen, F. Yang, H. Zheng, S. Feng, A proposal for low cross-talk square-lattice photonic crystal waveguide intersection utilizing the symmetry of waveguide modes, Optics communications, 273 (2007) 89-93.
[6] T. Liu, M. Fallahi, M. Mansuripur, A.R. Zakharian, J.V. Moloney, Intersection of nonidentical optical waveguides based on photonic crystals, Optics letters, 30 (2005) 2409-2411.
[7] K. Fasihi, S. Mohammadnejad, Orthogonal hybrid waveguides: an approach to low crosstalk and wideband photonic crystal intersections design, Journal of Lightwave Technology, 27 (2009) 799-805.
[8] J. Zhou, X. Di, D. Mu, J. Yang, W. Han, Improved broad bandwidth intersections in photonic crystals, Optics Communications, 285 (2012) 38-40.
[9] Y. Watanabe, Y. Sugimoto, N. Ikeda, N. Ozaki, A. Mizutani, Y. Takata, Y. Kitagawa, K. Asakawa, Broadband waveguide intersection with low-crosstalk in two-dimensional photonic crystal circuits by using topology optimization, Optics express, 14 (2006) 9502-9507.
[10] Y. Jiao, S.F. Mingaleev, M. Schillinger, D.A. Miller, S. Fan, K. Busch, Wannier basis design and optimization of a photonic crystal waveguide crossing, IEEE photonics technology letters, 17 (2005) 1875-1877.
[11] E. Danaee, A. Geravand, M. Danaie, Wide-band low cross-talk photonic crystal waveguide intersections using self-collimation phenomenon, Optics Communications, (2018).
[12] H. Chen, A.W. Poon, Low-loss multimode-interference-based crossings for silicon wire waveguides, IEEE photonics technology letters, 18 (2006) 2260-2262.
[13] C.-H. Chen, C.-H. Chiu, Taper-integrated multimode-interference based waveguide crossing design, IEEE Journal of Quantum Electronics, 46 (2010) 1656-1661.
[14] Y. Zhang, S. Yang, A.E.-J. Lim, G.-Q. Lo, C. Galland, T. Baehr-Jones, M. Hochberg, A CMOS-compatible, low-loss, and low-crosstalk silicon waveguide crossing, IEEE Photon. Technol. Lett, 25 (2013) 422-425.
[15] K.-S. Kim, Q.V. Vuong, Y. Kim, M.-S. Kwon, Compact Silicon Slot Waveguide Intersection Based on Mode Transformation and Multimode Interference, IEEE Photonics Journal, 9 (2017) 1-10.

[16] D. Chen, L. Wang, Y. Zhang, X. Hu, X. Xiao, S. Yu, Ultralow Crosstalk and Loss CMOS Compatible Silicon Waveguide Star-Crossings with Arbitrary Included Angles, ACS Photonics, 5 (2018) 4098-4103.
[17] W. Chang, L. Lu, X. Ren, D. Li, Z. Pan, M. Cheng, D. Liu, M. Zhang, Ultracompact dual-mode waveguide crossing based on subwavelength multimode-interference couplers, Photonics Research, 6 (2018) 660-665.
[18] H. Xu, Y. Shi, Dual-mode waveguide crossing utilizing taper-assisted multimode-interference couplers, Optics letters, 41 (2016) 5381-5384.
[19] W. Bogaerts, P. Dumon, D. Van Thourhout, R. Baets, Low-loss, low-cross-talk crossings for silicon-on-insulator nanophotonic waveguides, Optics letters, 32 (2007) 2801-2803.
[20] P. Sanchis, P. Villalba, F. Cuesta, A. Håkansson, A. Griol, J.V. Galán, A. Brimont, J. Martí, Highly efficient crossing structure for silicon-on-insulator waveguides, Optics letters, 34 (2009) 2760-2762.
[21] P.J. Bock, P. Cheben, J.H. Schmid, J. Lapointe, A. Delâge, D.-X. Xu, S. Janz, A. Densmore, T.J. Hall, Subwavelength grating crossings for silicon wire waveguides, Optics express, 18 (2010) 16146-16155.
[22] J. Feng, Q. Li, S. Fan, Compact and low cross-talk silicon-on-insulator crossing using a periodic dielectric waveguide, Optics letters, 35 (2010) 3904-3906.
[23] Y. Sakamaki, T. Saida, T. Hashimoto, S. Kamei, H. Takahashi, Loss reduction of waveguide crossings by wavefront matching method and their application to integrated optical circuits, Journal of Lightwave Technology, 27 (2009) 2257-2263.
[24] Y. Sakamaki, T. Saida, T. Hashimoto, H. Takahashi, New optical waveguide design based on wavefront matching method, Journal of Lightwave Technology, 25 (2007) 3511-3518.
[25] H.-L. Han, H. Li, X.-P. Zhang, A. Liu, T.-Y. Lin, Z. Chen, H.-B. Lv, M.-H. Lu, X.-P. Liu, Y.-F. Chen, High performance ultra-compact SOI waveguide crossing, Optics Express, 26 (2018) 25602-25610.
[26] L. Lu, M. Zhang, F. Zhou, W. Chang, J. Tang, D. Li, X. Ren, Z. Pan, M. Cheng, D. Liu, Inverse-designed ultra-compact star-crossings based on PhC-like subwavelength structures for optical intercross connect, Optics Express, 25 (2017) 18355-18364.
[27] Z. Yu, A. Feng, X. Xi, X. Sun, Inverse-designed low-loss and wideband polarization-insensitive silicon waveguide crossing, Optics Letters, 44 (2019) 77-80.
[28] W. Chang, L. Lu, X. Ren, L. Lu, M. Cheng, D. Liu, M. Zhang, An Ultracompact Multimode Waveguide Crossing Based on Subwavelength Asymmetric Y-Junction, IEEE Photonics Journal, 10 (2018) 1-8.
[29] M.M. Gilarlue, S.H. Badri, H. Rasooli Saghai, J. Nourinia, C. Ghobadi, Photonic crystal waveguide intersection design based on Maxwell's fish-eye lens, Photonics and Nanostructures - Fundamentals and Applications, 31 (2018) 154-159.
[30] H. Xu, Y. Shi, Metamaterial-Based Maxwell's Fisheye Lens for Multimode Waveguide Crossing, Laser & Photonics Reviews,  1800094.
[31] M.M. Gilarlue, J. Nourinia, C. Ghobadi, S.H. Badri, H.R. Saghai, Multilayered Maxwell's fisheye lens as waveguide crossing, Optics Communications, 435 (2019) 385-393.
[32] S. Li, Y. Zhou, J. Dong, X. Zhang, E. Cassan, J. Hou, C. Yang, S. Chen, D. Gao, H. Chen, Universal multimode waveguide crossing based on transformation optics, Optica, 5 (2018) 1549-1556.
[33] U. Leonhardt, T.G. Philbin, Transformation optics and the geometry of light,  Progress in Optics, Elsevier2009, pp. 69-152.
[34] L. Xu, H. Chen, Conformal transformation optics, Nature Photonics, 9 (2015) 15.
[35] M. McCall, J.B. Pendry, V. Galdi, Y. Lai, S. Horsley, J. Li, J. Zhu, R.C. Mitchell-Thomas, O. Quevedo-Teruel, P. Tassin, Roadmap on transformation optics, Journal of Optics, 20 (2018) 063001.
[36] O. Quevedo-Teruel, W. Tang, R.C. Mitchell-Thomas, A. Dyke, H. Dyke, L. Zhang, S. Haq, Y.J.S.r. Hao, Transformation optics for antennas: why limit the bandwidth with metamaterials?, 3 (2013) 1903.
[37] N. Kundtz, D.R.J.N.m. Smith, Extreme-angle broadband metamaterial lens, 9 (2010) 129.
[38] J. Li, J.B. Pendry, Hiding under the carpet: a new strategy for cloaking, Physical review letters, 101 (2008) 203901.

[39] B. Arigong, J. Shao, H. Ren, G. Zheng, J. Lutkenhaus, H. Kim, Y. Lin, H. Zhang, Reconfigurable surface plasmon polariton wave adapter designed by transformation optics, Optics express, 20 (2012) 13789-13797.
[40] B. Arigong, J. Ding, H. Ren, R. Zhou, H. Kim, Y. Lin, H. Zhang, Design of wide-angle broadband luneburg lens based optical couplers for plasmonic slot nano-waveguides, Journal of Applied Physics, 114 (2013) 144301.
[41] R. Duraiswami, A. Prosperetti, Orthogonal mapping in two dimensions, Journal of Computational Physics, 98 (1992) 254-268.
[42] S.H. Badri, M.M. Gilarlue, Maxwell's fisheye lens as efficient power coupler between dissimilar photonic crystal waveguides, Optik, 185 (2019) 566-570.
[43] Y. Dattner, O. Yadid-Pecht, Analysis of the effective refractive index of silicon waveguides through the constructive and destructive interference in a Mach–Zehnder interferometer, IEEE Photonics Journal, 3 (2011) 1123-1132.